\documentclass[aps,prc,twocolumn,superscriptaddress,nofootinbib,floatfix,longbibliography]{revtex4-2}

\usepackage{amsmath}
\usepackage{amssymb}
\usepackage{graphicx}
\usepackage{bm}
\usepackage{xcolor}
\usepackage{orcidlink}

\usepackage{hyperref}
\hypersetup{colorlinks=true,citecolor=blue,linkcolor=blue,urlcolor=blue}

\begin{document}

\title{Bridging \textit{Ab Initio} Symmetries and Global Nuclear Masses \\ with Interpretable Neural Networks}

\author{Phong Dang\,\orcidlink{0009-0002-7066-3988}}
\email{pdang5@lsu.edu}
\affiliation{Department of Physics and Astronomy, Louisiana State University, Baton Rouge, LA 70803, USA}

\author{Evander Espinoza}
\affiliation{Department of Physics and Astronomy, Louisiana State University, Baton Rouge, LA 70803, USA}

\author{Xiaoliang Wan\,\orcidlink{0009-0000-0146-1376}}
\affiliation{Department of Mathematics, Louisiana State University, Baton Rouge, LA 70803, USA}

\author{Michela Negro\,\orcidlink{0000-0002-6548-5622}}
\affiliation{Department of Physics and Astronomy, Louisiana State University, Baton Rouge, LA 70803, USA}

\author{Jerry P. Draayer\,\orcidlink{0000-0003-3568-8223}}
\affiliation{Department of Physics and Astronomy, Louisiana State University, Baton Rouge, LA 70803, USA}
\affiliation{Quantum CodeX, Baton Rouge, LA 70808, USA}

\author{Feng Pan\,\orcidlink{0000-0001-5118-1719}}
\affiliation{Department of Physics, School of Science, Huzhou Normal University, Huzhou 313000, China}
\affiliation{Department of Physics, Liaoning Normal University, Dalian 116029, China}
\affiliation{Department of Physics and Astronomy, Louisiana State University, Baton Rouge, LA 70803, USA}

\author{Tom\'{a}\v{s} Dytrych\,\orcidlink{0000-0002-1554-1462}}
\affiliation{Nuclear Physics Institute, Academy of Sciences of the Czech Republic, 25068 \v{R}e\v{z}, Czech Republic}
\affiliation{Department of Physics and Astronomy, Louisiana State University, Baton Rouge, LA 70803, USA}
\affiliation{Quantum CodeX, Baton Rouge, LA 70808, USA}

\author{Daniel Langr\,\orcidlink{0000-0001-9760-7068}}
\affiliation{Department of Computer Systems, Faculty of Information Technology, Czech Technical University, 16000 Prague, Czech Republic}
\affiliation{Quantum CodeX, Baton Rouge, LA 70808, USA}

\author{David Kekejian\,\orcidlink{0000-0001-8862-7681}}
\affiliation{Quantum CodeX, Baton Rouge, LA 70808, USA}

\begin{abstract}
\noindent
\textbf{Background:} \textit{Ab initio} theory has established Wigner's $\rm SU(4)$ and Elliott's $\rm SU(3)$ as dominant symmetries of the nuclear force in light and intermediate-mass nuclei. While previous work has established the relevance of the former symmetry for nuclear binding, whether the latter also organizes binding across the entire nuclear chart remains elusive.

\noindent
\textbf{Purpose:} We probe whether both the $\rm SU(3)$ and $\rm SU(4)$ symmetries organize nuclear masses, aiming at both physical insights through interpretable models and predictive capability.

\noindent
\textbf{Methods:} From a compact set of the $\rm SU(3)$ and $\rm SU(4)$ Casimir operators we build three neural-network mass models. Two are conventional---a feature-informed network (FINN) and a Gaussian variant (GINN) with predictive spread estimate---while the third is a new design: the Wigner-informed network (WINN) constrains the mass formula to be linear in the symmetry operators, learning only their $(N,Z)$-dependent couplings, so that the trained model is itself a transparent, intrinsically explainable mass relation. All are trained on \texttt{AME2016} subtracted by the liquid drop model at four data fractions and validated on nuclei new to \texttt{AME2020}, with extrapolation benchmarked against the HFB-26 model and $r$-process nucleosynthesis.

\noindent
\textbf{Results:} The Casimir features carry binding information far beyond the liquid-drop bulk, and a SHAP analysis suggests the quadratic $\rm SU(4)$ Casimir as the leading symmetry-derived contributor to the residual binding. The WINN yields the best performance---reaching a $0.412$~MeV validation error and competitive with state-of-the-art global models---and more importantly, even when trained on the sparsest dataset it still outperforms the other NNs trained on the densest.
Off the known chart its masses track HFB-26 as closely as WS3 and reproduce the solar abundance peaks of a neutron-star-merger $r$-process simulation. The WINN's coupling fields further reveal an enhanced even-$\rm SU(4)$ contribution toward the neutron dripline, hinting at a restoration of Wigner's symmetry.

\noindent
\textbf{Conclusions:} The $\rm SU(4)$- and $\rm SU(3)$-derived structures reach beyond individual nuclei to organize binding across the whole chart, and embedding symmetry-preserving operators directly in a domain-informed interpretable architecture yields a compact, physically transparent mass model that is less hungry for data.
\end{abstract}

\maketitle

\section{Introduction}
\label{intro}

Advances in computational hardware and data availability have catalyzed the adoption of machine learning (ML) across nuclear physics \cite{Boehnlein2022RMP}. Its reach spans a broad range of observables, including nuclear charge radii \cite{Utama2016JPG,Wu2020PRC,Meng2026PRC}, $\alpha$-decay \cite{Jalili2024SciRep} and $\beta$-decay \cite{Costiris2009PRC,Niu2019PRC} half-lives, and fission product yields \cite{Wang2019PRL}. In particular, neural networks (NNs) have shown strong performance in modeling nuclear mass---a central quantity for $r$-process nucleosynthesis \cite{Li2024PLB} and the composition of neutron star crusts \cite{Utama2016PRC}. Architectures explored to this end include light gradient boosting machine \cite{Gao2021NST}, mixture density network \cite{Mumpower2022PRC}, fully connected NNs \cite{Huang2025PRC,Kim2026PRC,Lu2026PRC}, decision trees \cite{Liu2025PRC}, convolutional NN \cite{Lu2025PRC}, Bayesian models \cite{Niu2018PLB,Utama2018PRC,Qu2025CPC,Kejzlar2023SciRep}, Gaussian processes and support vector regression \cite{Yuksel2024PRC,Ye2025PRC,Liu2026PRC}, as well as a hybrid between NN and density functional theory (DFT) \cite{Jalili2026JPG}. Additionally, symbolic regression has been employed to rediscover nuclear models from binding energy and charge radius data of light and medium mass nuclei \cite{Munoz2025CommPhys}.

The advantage of ML lies in its ability to capture recurring patterns that underpin physical systems. Atomic nuclei are particularly intriguing in this regard. Despite the complexity of the nuclear force, low-lying spectra are consistently dominated by two spatial symmetries, namely Elliott's $\rm SU(3)$ \cite{Elliott1958PRSA,Elliott1958PRSA-II,Elliott1963PRSA-III} and symplectic $\rm Sp(3,R)$ \cite{Rowe1985RPP} symmetries. Over the last two decades, \textit{ab initio} nuclear theory has unveiled the dominance of these symmetries in light and intermediate-mass nuclei \cite{Dytrych2007PRL,Dytrych2013PRL,Launey2016PPNP,Dytrych2020PRL,McCoy2020PRL}, leveraged them to model clustering phenomena and reactions \cite{Mercenne2022CPC,Launey2026PPNP}, and probed new physics beyond the Standard Model \cite{Sargsyan2022PRL,Launey2023SciPost}.

Prior to the introduction of the $\rm SU(3)$ symmetry, Wigner proposed the $\rm SU(4)$ [or $\rm U(4)$] symmetry in 1937 \cite{Wigner1937PR,Wigner1939PR}, for which he was awarded the Nobel Prize in 1963, alongside the pioneers of the nuclear shell model---Mayer \cite{Mayer1948PR,Mayer1949PR} and Jensen, Haxel and Suess \cite{Haxel1949PR}. Also known as the supermultiplet symmetry, it unifies the spin and isospin symmetries into a single framework. While the spin-orbit force breaking the symmetry in heavy nuclei cast doubt on the theory in the 1940s-50s, the 1960s saw renewed interest following the first observation of isobaric analog states in medium and heavy nuclei \cite{Anderson1962PR} and the formulation of a nuclear mass formula involving the second-order Casimir invariant of the $\rm SU(4)$ group as a test of this symmetry \cite{Franzini1963PL,Isacker1995PRL}. This model was later extended with higher-order Casimirs and shell corrections \cite{Cauvin1981NPA,Gaponov1982NPA,Isacker1997FoP}, exceeding the Bethe-Weizs\"acker formula \cite{Weizsacker1935ZTK}, which treats nuclei as liquid drops (LDM).

Approaching its nine-decade milestone, Wigner's symmetry has remained at the forefront of nuclear physics. First-principles studies have recently established $\rm SU(4)$ as a dominant symmetry of the nuclear force, which arises naturally from the strong interaction in effective field theories \cite{Kaplan1996PLB,Kaplan1997PRC,Cordon2008PRC,Lee2021PRL,Beane2019PRL}, is strongly manifested in light nuclei \cite{Muli2026PRL,Dang2026}, and governs the coupling of nuclear structure to the weak force \cite{Muli2026PRL}. These insights have been exploited to tame the sign problem of nuclear lattice simulations \cite{Lee2025ARNPS,Niu2025PRL} and to compress no-core shell-model spaces via the \textit{ab initio} Symmetry Adapted Model (SAM) \cite{Dang2026}. As with Elliott's $\rm SU(3)$, however, this evidence has been gathered mostly in light and intermediate-mass nuclei, where computations of nuclei from first principles are  tractable on leadership-class supercomputers. 

Meanwhile, tremendous theoretical and computational development has been made to extend the reach of \textit{ab initio} calculations well into the region of medium-mass and heavy nuclei \cite{Hergert2020FP}. Such predictive calculations are of high importance for modeling and understanding the formation of heavy elements in the universe through rapid neutron captures that occur in explosive astrophysical environments which will be surveyed by NASA's Compton Spectrometer and Imager (COSI) mission \cite{Tomsick2024ICRC}, planned for launch in 2027. This nucleosynthesis process is known as the $r$-process, producing nuclei along extremely neutron-rich paths. While efforts at frontier facilities like FRIB, TRIUMF, GSI, GANIL are expanding the nuclear landscape beyond the last measured isotopes, information about many $r$-process nuclei is unknown and has been inferred from global phenomenological models. Only recently has \textit{ab initio} theory begun to provide input for the $r$-process-relevant region, with masses and their uncertainties computed near the $N=82$ shell closure \cite{Kuske2026PRL} through an \textit{ab initio} many-body method called the valence-space in-medium similarity renormalization group \cite{Stroberg2019ARNPS,Stroberg2021PRL}.

In this paper, we present the first NN model (to our knowledge) of nuclear binding energies explicitly informed by the $\rm SU(3)$ and $\rm SU(4)$ symmetries. Traditionally, these emergent symmetries have guided the construction of the many-body basis of individual nuclei. Here we broaden their reach by turning the same symmetries, grounded in first principles, into a lens for probing and inferring nuclear systematics across the entire chart. Our novel contribution is three-fold. 

First, from a compact set of features built on the $\rm SU(3)$ and $\rm SU(4)$ Casimir invariants we construct two black-box NN mass models---FINN and GINN---and establish that these symmetry features carry binding-energy information well beyond the bulk properties provided by the LDM. 

Second, and central to this work, we introduce WINN, an NN mass model that embeds the symmetry features directly through a formula linear in the operators rather than through a direct black-box NN. While Ref. \cite{Kim2026PRC} incorporates domain knowledge through the activation function, the WINN incorporates it structurally through the symmetry-operator basis of the mass formula, constraining the output to remain linear in the operators while allowing their $(N,Z)$-dependent couplings to be learned. The WINN attains a validation root-mean-square error (RMSE) of $0.412$~MeV, competitive with state-of-the-art mass models and lower than the less transparent FINN and GINN. It is also markedly more data-efficient---trained on only a fifth of the data, the WINN already outperforms the FINN and GINN trained on the largest set. We also find that the WINN exhibits more stable extrapolation of single-neutron separation energy---an essential quantity for the $r$-process---for the Sn and Pb isotopic chains compared to the FINN and GINN. Together, these provide evidence that building the symmetry map into the architecture is more effective than supplying it as input features alone.

Third, the WINN is itself intrinsically interpretable, shedding light on the role of the symmetries for various mass regions without any feature-attribution techniques. Specifically, only the WINN's $(N,Z)$-dependent coupling strengths are learned, so the trained model is a transparent mass relation whose contributions can be read term by term across the chart. This minimal symmetry basis invites broader extensions in both symmetry content and the range of observables for future work.

The remainder of this paper is organized as follows. Section \ref{methods} describes the datasets, the symmetry-inspired features, the three NN architectures and the training strategy. Section \ref{results} reports the accuracy of the models, interprets the learned physics through a SHAP analysis and the WINN coupling strengths, and benchmarks the resulting mass tables via $r$-process nucleosynthesis in the dynamic ejecta of a neutron star merger. Key findings and future directions are summarized in Sec.~\ref{summary}.

\begin{figure*}[!tbp]
  \centering
  \includegraphics[width=0.9\textwidth]{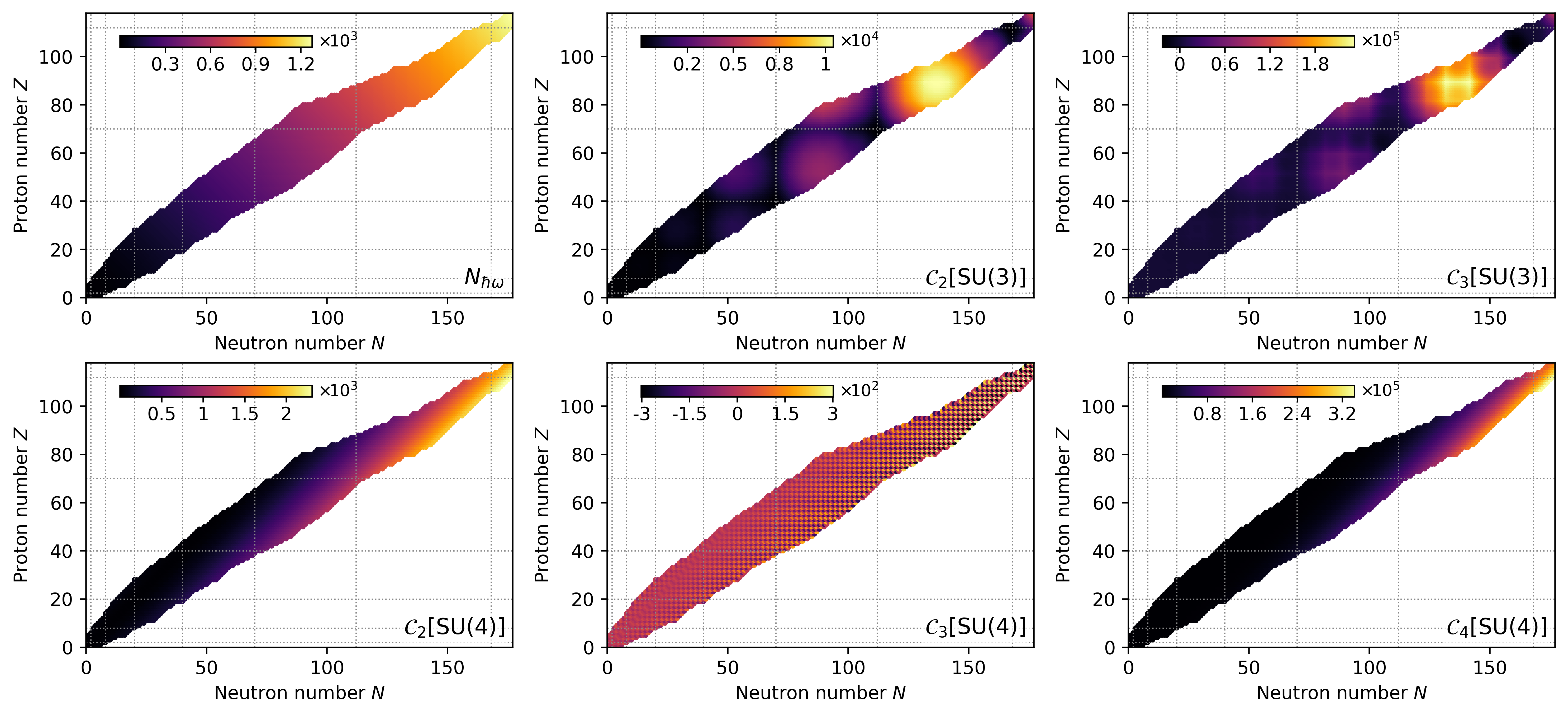}
  \caption{Color maps that show evolution of six operators, including total harmonic oscillator quanta ($N_{\hbar\omega}$), $\rm SU(3)$ Casimir operators ($\mathcal{C}_2$, $\mathcal{C}_3$), $\rm SU(4)$ Casimir operators ($\mathcal{C}_2$, $\mathcal{C}_3$, $\mathcal{C}_4$), across the entire nuclear chart. The dotted lines correspond to closed harmonic oscillator shells for protons and neutrons.}
  \label{fig:features}
\end{figure*}

\section{Methods}
\label{methods}

\subsection{Datasets and targets}
\label{datasets}

We collect experimental nuclear masses from \texttt{AME2016} \cite{AME2016_1,AME2016_2} and \texttt{AME2020} \cite{AME2020_1,AME2020_2}, excluding values extrapolated from systematics and light nuclei with $Z<8$ or $N<8$. Nuclei measured in \texttt{AME2016} form the train/test pool, while those first reported in \texttt{AME2020} (71 nuclei) provide an unseen validation set.

Since binding energies are relatively smooth functions of $N$ and $Z$, a large train set guarantees that a randomly held-out nucleus for testing is almost invariably surrounded by training neighbors, thus its accurate prediction reflects little more than interpolation between adjacent data. Similarly, the newly measured nuclei in \texttt{AME2020} sit only one or two neutrons away from the training set. Therefore, in order to put our models to a more stringent test, we pursue two strategies.

First, inspired by Refs. \cite{Mumpower2022PRC,Li2024PLB,Yuksel2024PRC}, we train each model at four training fractions rather than a single split, $f \in \{0.8,\,0.6,\,0.4,\,0.2\}$, of the \texttt{AME2016} pool, retaining the complement as a test set that is drawn from the same distribution as the train set. As $f$ decreases, the training data thin out and the gaps around the held-out nuclei widen; as a result, a model that merely interpolates between neighbors deteriorates rapidly, whereas one that has grasped the underlying systematics should hold steady. 

Second, and more importantly, we confront the models with extrapolation into unexplored territory. For every element, we identify its last measured isotope and evaluate the models on neutron-rich nuclei up to 40 neutrons beyond it ($0 < d \le 40$, with $d$ as the neutron distance from the measured edge).
As experimental masses are unavailable in this region, we gauge the predictions against an established global mass model---the fully microscopic HFB-26 \cite{Goriely2013PRC}.

We note that extrapolating a data-driven model far beyond its training domain is inherently perilous on the ground that with no data off stability, the networks are free to wander toward unphysical values. To curb this undesired behavior, we follow the strategy shown in Ref. \cite{Li2024PLB} and augment the train set with 50 nuclei drawn from the macroscopic-microscopic WS3 model \cite{Wang2010PRC} in the extrapolation region, stratified across the distance $d$ of $0$-$5$, $5$-$10$, $10$-$20$, and $20$-$40$ neutrons with 10, 13, 15, and 12 isotopes, respectively. Sparse as they are, these anchors cannot fit the unmeasured region consisting of more than 7000 nuclei predicted to exist; rather, they nudge our NN mass models toward physically sensible behavior beyond the currently known edge, appreciably stabilizing the predictions there while leaving the interpolation performance unchanged.

Furthermore, following Refs. \cite{Utama2016PRC,Gao2021NST,Huang2025PRC}, we remove a macroscopic baseline and let the networks focus on the fine structure of binding energy ($BE$) by subtracting the five-term LDM \cite{Weizsacker1935ZTK},
\begin{align}
\label{eq:ldm}
BE_{\rm LDM} = a_V A &- a_S A^{2/3} - a_C \frac{Z(Z-1)}{A^{1/3}} \nonumber \\
&- a_A \frac{(N-Z)^2}{A} + a_P \frac{\delta}{\sqrt{A}},
\end{align}
with $\delta = +1,\,-1,\,0$ for even-even, odd-odd, and odd-$A$ nuclei. Least-squares fitting of Eq.~(\ref{eq:ldm}) to the known masses in \texttt{AME2016} yields $a_V = 15.58$~MeV, $a_S = 17.39$~MeV, $a_C = 0.706$~MeV, $a_A = 22.91$~MeV, and $a_P = 12.52$~MeV. This baseline is held fixed across all models and training fractions, and each network learns only the residual $\Delta BE = BE - BE_{\rm LDM}$. Subtracting the common bulk baseline isolates the residual binding structure on which the three architectures are compared.

\subsection{Symmetry-inspired features}
\label{features}

We employ features grouped into two categories. 

\textit{SU(4) features}---Inspired by earlier mass models using Wigner's $\rm SU(4)$ symmetry \cite{Cauvin1981NPA,Gaponov1982NPA,Isacker1997FoP}, we leverage the Casimir operators of the $\rm SU(4)$ group to be features. These include the two-, three-, and four-body operators, $\mathcal{C}_2$, $\mathcal{C}_3$, $\mathcal{C}_4$, which describe spin-isospin polarization forces. (Their eigenvalues are derived in Appendix \ref{app:su4}.) Since these operators depend only on $\rm SU(4)$ irreducible representations (irreps) and several irreps are allowed for a given nucleus, we hypothesize that the preference toward the lowest-$\mathcal{C}_2$ irrep revealed by first-principles calculations for light nuclei \cite{Muli2026PRL,Dang2026} also extends to the entire nuclear chart. Figure \ref{fig:features} shows evolution of these operators. While the quadratic and quartic operators gradually increase for heavier nuclei, the cubic Casimir vanishes for every even-mass nucleus and exhibits a sign fluctuation between odd-mass nuclei depending on whether a proton or a neutron is unpaired, which becomes increasingly pronounced for very neutron-rich isotopes.

\textit{SU(3) features}---We select three features stemming from the three-dimensional harmonic oscillator (3DHO) shells. One is the number operator $N_{\hbar\omega}$ that counts the total number of oscillator quanta of a particular distribution of nucleons across the shells. The others are the second- and third-order invariants of the $\rm SU(3)$ group, whose eigenvalues are given in Appendix~\ref{app:su3}. The former, $\mathcal{C}_2$, signifies the magnitude of the mass quadrupole moment, whereas the latter, $\mathcal{C}_3$, indicates whether the nucleus is spherical (zero), prolate (positive), or oblate (negative) \cite{Draayer1989PRL,Castanos1988ZPA}. For each nucleus, an $\rm SU(3)$ irrep is selected such that it combines antisymmetrically with the $\rm SU(4)$ irrep and maximizes deformation \cite{Dytrych2013PRL}. Across the chart, these operators peak at mid-shells where nuclei tend to deform and get suppressed at shell closures where nuclei are mostly spherical (Fig. \ref{fig:features}).

\begin{figure}[t!]
  \centering
  \includegraphics[width=\linewidth]{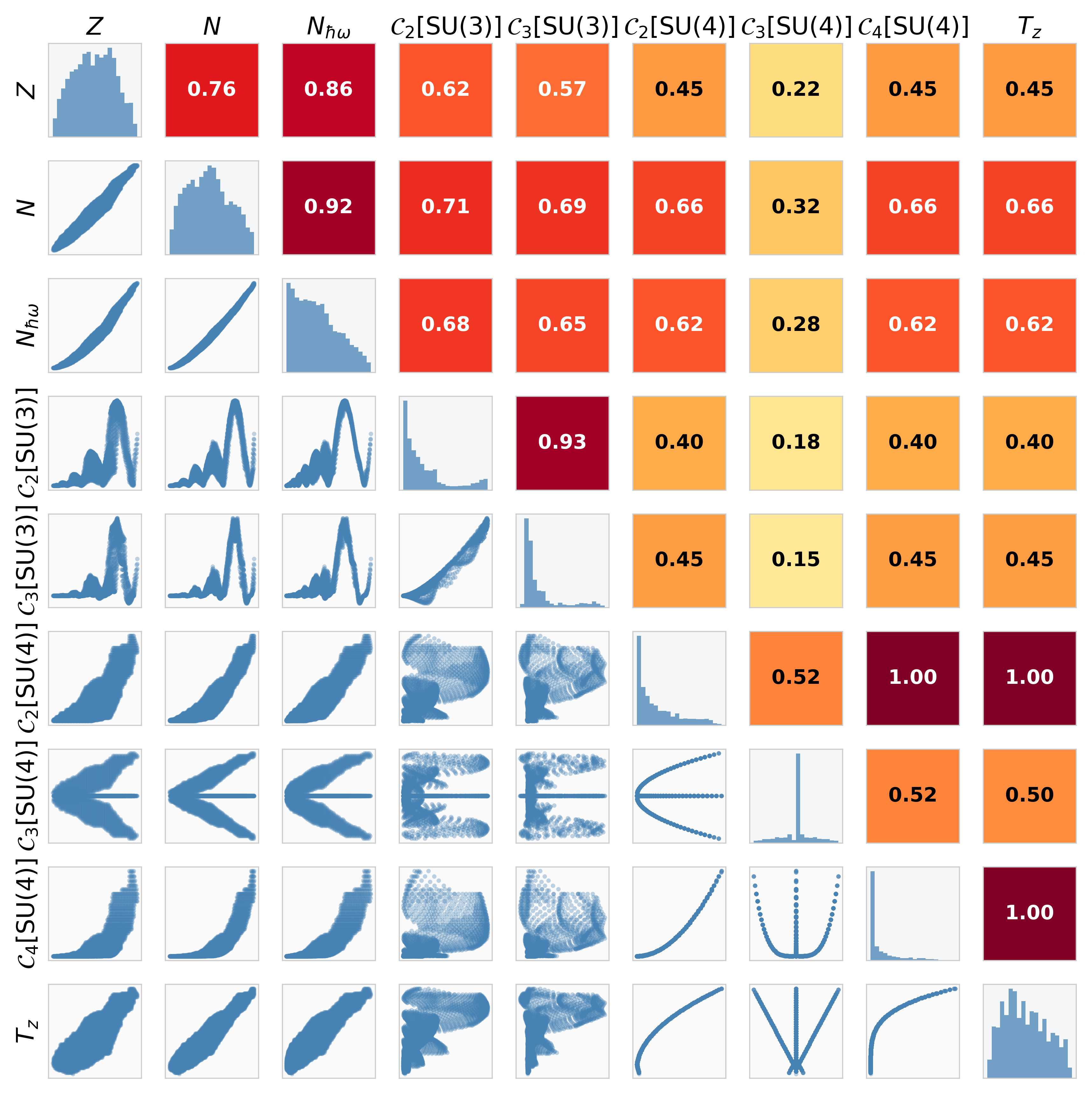}
  \caption{Association matrix of nine quantities, namely proton number ($Z$), neutron number ($N$), total harmonic oscillator quanta ($N_{\hbar\omega}$), $\rm SU(3)$ Casimir operators ($\mathcal{C}_2$, $\mathcal{C}_3$), $\rm SU(4)$ Casimir operators ($\mathcal{C}_2$, $\mathcal{C}_3$, $\mathcal{C}_4$), and isospin projection ($T_z = (N-Z)/2$). Along the diagonal are distributions of the quantities; the upper half of the matrix contains the maximal information coefficients, whereas the lower half displays scatter plots for all pairs.}
  \label{fig:correlations}
\end{figure}

\begin{figure*}[t!]
    \centering
    \includegraphics[width=\textwidth]{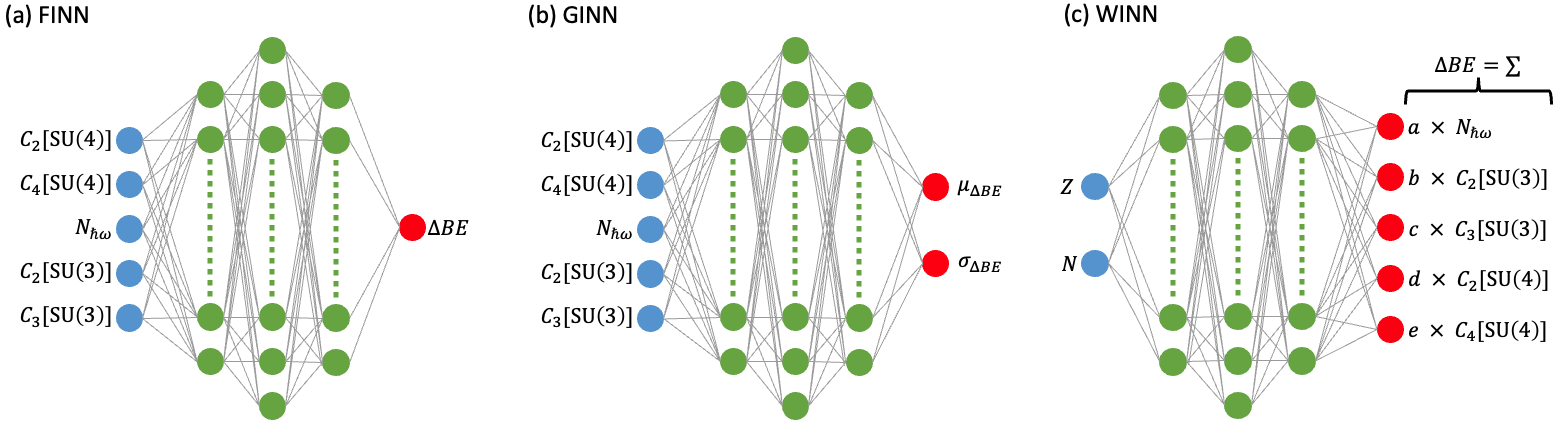}
    \caption{Neural network architectures for (a) FINN, (b) GINN, and (c) WINN. Blue nodes are input features; green nodes are hidden neurons arranged in three layers of 64, 128, and 64 units; red nodes are outputs. (a) FINN produces a single point prediction $\Delta BE$. (b) GINN shares the same backbone as FINN but branches into two independent linear heads, yielding a predicted mean $\mu_{\Delta BE}$ and a spread $\sigma_{\Delta BE}$. (c) WINN admits only $(Z,N)$ as inputs and outputs $(N,Z)$-dependent coupling strengths; the binding energy difference $\Delta BE$ is the sum of each coupling multiplied by its symmetry feature. FINN and WINN use ReLU activation function, while GINN uses its parametric variant (PReLU).}
    \label{fig:architectures}
\end{figure*}

\textit{Feature associations}---Before training our models, we examine the pairwise relationships between the features and $N$, $Z$, $T_z$. As the scatter plots in Fig.~\ref{fig:correlations} show, these relationships are predominantly non-monotonic; hence, we quantify them with the maximal information coefficient (MIC) \cite{Reshef2011Science}, which captures both linear and nonlinear dependence, rather than the Pearson or Spearman coefficients \cite{Schober2018AnesAnal}. 
First, $N_{\hbar\omega}$ is strongly associated with $N$ and $Z$ (MIC $\approx 0.9$), indicating that it largely tracks the mass number $A$, consistent with its evolution in Fig.~\ref{fig:features}. 
Second, the $\rm SU(3)$ invariants are tightly coupled to each other (MIC $=0.93$), as both encode quadrupole collectivity, while moderately associated with the bulk features (MIC $\approx 0.6$-$0.7$). 
Third, the even $\rm SU(4)$ Casimirs ($\mathcal{C}_2$ and $\mathcal{C}_4$) exhibit a deterministic functional relationship with $T_z$ (MIC $=1.00$), which means, for the dominant irrep, their eigenvalues reduce to smooth even functions of $T_z$. Since these operators are non-negative and quadratic and quartic in degree, we fit their eigenvalues against polynomials of $|T_z|$, yielding
\begin{equation} 
\langle \mathcal{C}_2 [\rm SU(4)] \rangle \simeq 2.00 |T_z|^2 + 7.97 |T_z| + 3.20
\end{equation}
with $R^2 = 0.99998$ and 
\begin{align}
\langle \mathcal{C}_4 [\rm SU(4)] \rangle \simeq 0.25 |T_z|^4 &+ 1.90 |T_z|^3 + 11.9 |T_z|^2 \nonumber \\
&+ 15.5 |T_z| + 22.9
\end{align}
with $R^2 = 0.99991$, where $R^2$ is the coefficient of determination. Crucially, the linear $|T_z| \propto |N-Z|$ term in both $\mathcal{C}_2 [\rm SU(4)]$ and $\mathcal{C}_4 [\rm SU(4)]$ represents the Wigner energy \cite{Satula1997PLB}, which is known as the cusp at $N=Z$. Therefore, the inclusion of $\mathcal{C}_2 [\rm SU(4)]$ on top of the asymmetry term ($\propto T_z^2$) in the LDM is equivalent to adding the Wigner energy, whereas $\mathcal{C}_4 [\rm SU(4)]$ brings in higher orders of isospin asymmetry that embody the supermultiplet structure as a physically motivated inductive bias. These polynomial relationships apply only to the leading irrep that is constrained directly by the proton and neutron numbers. Subleading irreps need not be constrained by $N$ and $Z$ in the same way, and their Casimir eigenvalues may therefore no longer be smooth functions of $T_z$, which we will investigate in future work.
Fourth, $\mathcal{C}_3[\rm SU(4)]$ is weakly associated with every other feature (MIC $\leq 0.32$ with the bulk and $\rm SU(3)$ sectors\footnote{The Pearson and Spearman coefficients are zero for these relationships, showing that they are unable to detect these complex associations, despite the visible trends in the scatter plots.}), reflecting the sign fluctuation noted above. 
Given its weak association with the remaining features and rapid sign fluctuations, we omit $C_3[SU(4)]$ from the present work.
The final feature set therefore comprises the five operators, namely $N_{\hbar\omega},\, \mathcal{C}_2[\rm SU(3)],\, \mathcal{C}_3[\rm SU(3)],\, \mathcal{C}_2[\rm SU(4)],\, \mathcal{C}_4[\rm SU(4)]$.

\subsection{Neural-network architectures}
\label{arch}

From the symmetry features we build three NNs that share a backbone of three hidden layers (64, 128, 64 units) but differ in how they map features to the binding-energy residual $\Delta BE$ (Fig.~\ref{fig:architectures}).

\textit{FINN} (Feature-Informed NN) is a standard fully-connected multilayer perceptron that takes the full feature vector and returns a single residual $\Delta BE$, thus a point estimate of $BE$, see Fig.~\ref{fig:architectures}(a).

\textit{GINN} (Gaussian-Informed NN) shares the FINN's backbone but splits into two heads after the last hidden layer, returning a mean $\mu_{\Delta BE}$ and a predictive spread $\sigma_{\Delta BE}$ as illustrated in Fig.~\ref{fig:architectures}(b), which is equivalent to the mixture density network with one Gaussian previously applied to nuclear masses in Ref. \cite{Mumpower2022PRC}. The goal of this architecture is to supply a per-nucleus predictive spread that the FINN cannot. 

\textit{WINN} (Wigner-Informed NN) is the centerpiece of this work. While the FINN and GINN reprise fully-connected and mixture-density architectures that have already been established for nuclear masses in previous works \cite{Mumpower2022PRC,Huang2025PRC}, the WINN is a new design. Specifically, the other two architectures learn a direct mapping, $\mathbf{y} = \mathcal{F}(\mathbf{x})$, from features to the binding energy difference, which generically entangles the inputs in a nonlinear manner, thereby leaving obscure how each symmetry operator shapes binding across the chart. Motivated by earlier algebraic mass formulas \cite{Cauvin1981NPA,Isacker1997FoP}, we instead constrain the functional mapping to be linear in terms of the symmetry operators,
\begin{align}
\label{eq:winn}
\Delta BE = a\,N_{\hbar\omega} & + b\,\mathcal{C}_2[{\rm SU(3)}] + c\,\mathcal{C}_3[{\rm SU(3)}] \nonumber\\
& + d\,\mathcal{C}_2[{\rm SU(4)}] + e\,\mathcal{C}_4[{\rm SU(4)}],
\end{align}
where the coupling strengths $\{a,b,c,d,e\}$ are learned by the NN as functions of $(N,Z)$ as shown in Fig.~\ref{fig:architectures}(c). Because the operators enter strictly linearly, the trained WINN is itself an interpretable mass formula, i.e., the contribution of each term can be read off across the entire nuclear chart (see Sec.~\ref{winn-interp}), laying bare where and how each symmetry contributes to the learned binding systematics and providing a transparency that a black-box regressor cannot offer.

\begin{table*}[t!]
\caption{Root-mean-square error (RMSE, in MeV) of FINN, GINN, and WINN on the training, held-out test (\texttt{AME2016}), and validation (\texttt{AME2020}) sets at four training fractions $f$ of \texttt{AME2016}. For comparison, on the same validation set (the 71 nuclei new to \texttt{AME2020}, unseen by all models) the LDM baseline, WS3, and HFB-26 give RMSEs of $3.14$, $0.56$, and $0.78$~MeV, respectively.}
\label{tab:rmse}
\begin{ruledtabular}
\begin{tabular}{cccccccccc}
 & \multicolumn{3}{c}{FINN} & \multicolumn{3}{c}{GINN} & \multicolumn{3}{c}{WINN}\\
\cline{2-4}\cline{5-7}\cline{8-10}
$f$ & Train & Test & Val. & Train & Test & Val. & Train & Test & Val.\\
\colrule
0.8 & 0.371 & 0.469 & 0.856 & 0.360 & 0.459 & 0.831 & 0.287 & 0.412 & 0.412\\
0.6 & 0.366 & 0.451 & 0.930 & 0.322 & 0.456 & 0.900 & 0.250 & 0.368 & 0.516\\
0.4 & 0.344 & 0.547 & 1.034 & 0.333 & 0.526 & 0.986 & 0.270 & 0.413 & 0.598\\
0.2 & 0.436 & 0.896 & 1.751 & 0.492 & 0.882 & 1.429 & 0.381 & 0.713 & 0.774\\
\end{tabular}
\end{ruledtabular}
\end{table*}

All three networks are trained on the LDM residual including the 50 WS3 anchors as described in Sec.~\ref{datasets} with the \texttt{SOAP} optimizer \cite{Vyas2025} and a physics-informed loss $\mathcal{L} = \mathcal{L}_0 + \lambda\,\mathcal{L}_1$. Here, $\mathcal{L}_0$ is an architecture-based loss, i.e., the mean-squared error (MSE) for the FINN and WINN, and the negative log-likelihood (NLL) loss [$\langle (\Delta BE - \mu_{\Delta BE})^2/2\sigma_{\Delta BE}^2 + \log \sigma_{\Delta BE} \rangle$] for the GINN, while $\mathcal{L}_1$ is a physics-informed loss computed on the training set and multiplied by the weight $\lambda$ adapted from Refs. \cite{Barea2008PRC,Mumpower2022PRC,Yuksel2024PRC} and based on the Garvey-Kelson (GK) relations \cite{Garvey1969RMP},
\begin{align}
\label{GK1}
    & BE(Z-2,N+2) + BE(Z-1,N) \nonumber \\
    & + BE(Z,N+1) - BE(Z-2,N+1) \nonumber \\
    &  - BE(Z-1,N+2) - BE(Z,N) \approx 0
\end{align}
for $N \ge Z$ and
\begin{align}
\label{GK2}
    & BE(Z+2,N-2) + BE(Z+1,N) \nonumber \\
    & + BE(Z,N-1) - BE(Z+2,N-1) \nonumber \\ 
    & - BE(Z+1,N-2) - BE(Z,N) \approx 0
\end{align}
for $N < Z$.
While tuning the hyperparameters for the best performance on \texttt{AME2016}, we find that the GK weight differs sharply between the NNs; specifically, $\lambda = 1.0$ for the FINN and GINN but only $\lambda = 0.05$ for WINN, which is a factor of twenty smaller. This reduced dependence on the GK term is consistent with an architectural advantage of the WINN. Indeed, the FINN and GINN have no built-in knowledge of how $\Delta BE$ scales with $N$ and $Z$, so the GK term is essential to enforce local consistency across the chart. For the WINN, by contrast, the operator basis of Eq.~(\ref{eq:winn}) already encodes this regularity, so the GK term serves as a lightweight refinement rather than a structural constraint---an advantage that stems from building the symmetry systematics directly into the architecture from the outset, rather than feeding them simply as inputs.

There is a structural disadvantage of the GINN compared with the FINN. In the NLL loss the gradient on $\mu_{\Delta BE}$ is weighted by $1/\sigma_{\Delta BE}^2$, thus the network can lower the loss by inflating $\sigma_{\Delta BE}$, trading point accuracy for wider intervals. Meanwhile, the FINN, minimizing the MSE directly, is immune to this trade-off and tends to retain an RMSE edge. To mitigate this issue, we adopt the $\beta$-NLL weighting \cite{Seitzer2022} that rescales each loss by a stop-gradient factor $\sigma_{\Delta BE}^{2\beta}$ with $\beta = 0.5$ to tighten point accuracy. Additionally, the FINN and WINN are trained for 1000 epochs with weight decay of $10^{-2}$, whereas the GINN is trained for 1500 epochs without weight decay. All models use mini-batches of 64 and a learning rate of $10^{-3}$. 

\subsection{Feature normalization}

The FINN and GINN ingest the five operators directly. As Fig.~\ref{fig:features} shows, these operators span widely different scales, with the harmonic oscillator quanta $N_{\hbar\omega}$ ranging over the hundreds and the $\rm SU(4)$ and $\rm SU(3)$ Casimir eigenvalues growing even faster with nucleon number. Feeding such heterogeneous magnitudes directly into an NN is detrimental, since features with large numerical values dominate the gradients irrespective of their physical relevance, the optimization becomes ill-conditioned, and the inputs are pushed into the saturated regions of the activation functions. Hence, in order to balance the gradients, stabilize convergence, and ensure that the relative importance the models assign to each feature reflects its predictive value, we standardize every input feature and the target to zero mean and unit variance through $z$-score normalization,
\begin{equation}
\tilde{x}_{ij} = \frac{x_{ij} - \mu_i}{\sigma_i + \epsilon}, \qquad
\tilde{y}_j = \frac{y_j - \mu_y}{\sigma_y + \epsilon},
\label{eq:zscore}
\end{equation}
where $(\mu_i, \sigma_i)$ are the mean and standard deviation of feature $i$, $(\mu_y, \sigma_y)$ are those of the residual binding energy, and index $j$ represents a nucleus. Here, $\epsilon = 10^{-8}$ is a small constant that guards against division by zero. Crucially, the statistics $(\mu, \sigma)$ are computed on the train set only and then applied to the test and validation sets. This prevents information of the held-out nuclei from leaking into the preprocessing, so that the test and validation errors remain honest measures of generalization. Since the target is also standardized, the network learns in normalized units, and all predictions are mapped back onto physical units (MeV) through the inverse transform $y_j = \tilde{y}_j \sigma_y + \mu_y$ before the RMSE is reported.

The WINN handles normalization differently. Its network ingests only $(N,Z)$, which are comparable; these two inputs and the target are standardized by Eq.~(\ref{eq:zscore}) as above, while the operators enter as the multiplicative basis of Eq.~(\ref{eq:winn}) and never reach the activation functions where their disparate scales would do harm. The operators are therefore rescaled by their training-set standard deviation alone and are deliberately left un-centered, since subtracting a mean would spoil the term-by-term reading that underpins the WINN's interpretability. 

\section{Results and interpretations}
\label{results}

\subsection{Prediction accuracy}
\label{rmse}

Table~\ref{tab:rmse} summarizes the performance of the three models across the four training fractions. All of them generalize well, the WINN and GINN reproduce the newly measured \texttt{AME2020} nuclei to better than $1$~MeV down to $f=0.4$, while the FINN remains close at $1.03$~MeV. 
Among them the WINN stands out as at $f=0.8$ it attains a validation RMSE of $0.412$~MeV, competitive with modern global mass models, and it degrades far more slowly than the black-box FINN and GINN as data are withdrawn, retaining a $0.774$~MeV validation error even at the most sparse fraction $f=0.2$, that is, with only a fifth of \texttt{AME2016}. Strikingly, upon validation the WINN at $f=0.2$ already surpasses the FINN (0.856 MeV) and GINN (0.831 MeV) at $f=0.8$. Moreover, the WINN yields validation RMSEs that are roughly half of those from the other NNs across all training fractions. We interpret this data efficiency as evidence for the strong inductive bias imposed by the WINN: indeed, the form of Eq.~(\ref{eq:winn})---linear in the symmetry operators with $(N,Z)$-dependent coefficients---constrains the mass surface far more tightly than the unconstrained FINN and GINN maps, leaving little room to overfit sparse training data. Importantly, since all three models are built on the same symmetry operators---the FINN and GINN receive them as inputs, whereas the WINN takes $(N,Z)$ and applies its learned couplings to the same operators---the WINN's edge over the others supports the advantage of imposing the symmetry-operator basis as a linear architectural constraint rather than presenting it as inputs to a black box.

\begin{table}[t!]
\caption{Extrapolation mean absolute deviation (MAD, in MeV) between each NN model and HFB-26 over the unmeasured region within the distance of $0 < d \le 40$ neutrons from the last known isotopes, at four training fractions. For reference, the MAD between the two established models WS3 and HFB-26 over the same set is $5.53$~MeV.}
\label{tab:extrap}
\begin{ruledtabular}
\begin{tabular}{cccc}
$f$ & FINN & GINN & WINN\\
\colrule
0.8 & 5.40 & 5.63 & 5.05\\
0.6 & 4.98 & 6.38 & 4.70\\
0.4 & 4.95 & 4.98 & 4.95\\
0.2 & 5.47 & 5.29 & 5.37\\
\end{tabular}
\end{ruledtabular}
\end{table}

\begin{table}[t!]
\caption{Mean predicted $\sigma_{\Delta BE}$ of GINN (in MeV) versus the neutron distance $d$ beyond the last measured isotope, at training fractions $f=0.4$-$0.8$. The predicted spread grows with distance into the unmeasured region, flagging where the mass predictions become less confident.}
\label{tab:ginn_sigma}
\begin{ruledtabular}
\begin{tabular}{lccc}
$d$ (neutrons) & $f=0.8$ & $f=0.6$ & $f=0.4$ \\
\colrule
$\leq 0$    & 0.20 & 0.18 & 0.20 \\
$0$-$5$     & 0.21 & 0.27 & 0.17 \\
$5$-$10$    & 0.25 & 0.37 & 0.20 \\
$10$-$20$   & 0.38 & 0.34 & 0.25 \\
$20$-$40$   & 3.1  & 2.6  & 3.2  \\
\end{tabular}
\end{ruledtabular}
\end{table}

To probe extrapolation, we examine the mean absolute deviation (MAD) between each NN model and the HFB-26 mass table in the unmeasured region. We deliberately benchmark against HFB-26 rather than WS3, since the networks are anchored to the latter and such a comparison would be circular. Table~\ref{tab:extrap} shows that all three models follow HFB-26 to within about $5$~MeV at every training fraction---comparable to the $5.53$~MeV MAD that separates the two established models WS3 and HFB-26 themselves. One can say that in terra incognita, our NNs are as faithful to HFB-26 as a purpose-built global model is, out to 40 neutrons beyond the last known nuclei and with only a mild dependence on the training fraction. We stress, however, that because the networks are tied to WS3, this level of agreement with HFB-26 largely reflects the WS3 anchoring---in fact, the models differ from WS3 only ${\sim}1$~MeV---so the MAD does not by itself discriminate between the three architectures; the sharper, model-distinguishing extrapolation test is the $S_n$ systematics that is discussed below in Subsec. \ref{rprocess}.

The GINN, beyond its point predictions, supplies a per-nucleus $\sigma_{\Delta BE}$ that acts as a data-driven confidence indicator. The predicted $\sigma_{\Delta BE}$ is not interpreted as a calibrated uncertainty; rather, we use it as a relative indicator of where the GINN becomes less confident as it moves away from measured nuclei. Table~\ref{tab:ginn_sigma} reports the mean of $\sigma_{\Delta BE}$ across the chart as a function of the neutron distance $d$ from the last measured isotopes. We restrict the comparison to $f\ge0.4$, since at $f=0.2$ the GINN's validation accuracy itself degrades sharply (Table~\ref{tab:rmse}), leaving its predicted spread unreliable. Near measured nuclei the predicted $\sigma_{\Delta BE}$ remains at the ${\sim}0.2$~MeV level and grows with the extrapolation distance, by an order of magnitude to ${\sim}3$~MeV in the $d=20$-$40$ region. As the model departs further from the known nuclear chart, where data grow sparse, it becomes correspondingly less confident and registers this directly through a rising $\sigma_{\Delta BE}$. The GINN thus carries a built-in, distance-aware flag for the model confidence of each predicted mass. 

\subsection{Feature importance}
\label{shap}

\begin{figure}[h!]
    \centering
    \includegraphics[width=0.48\textwidth]{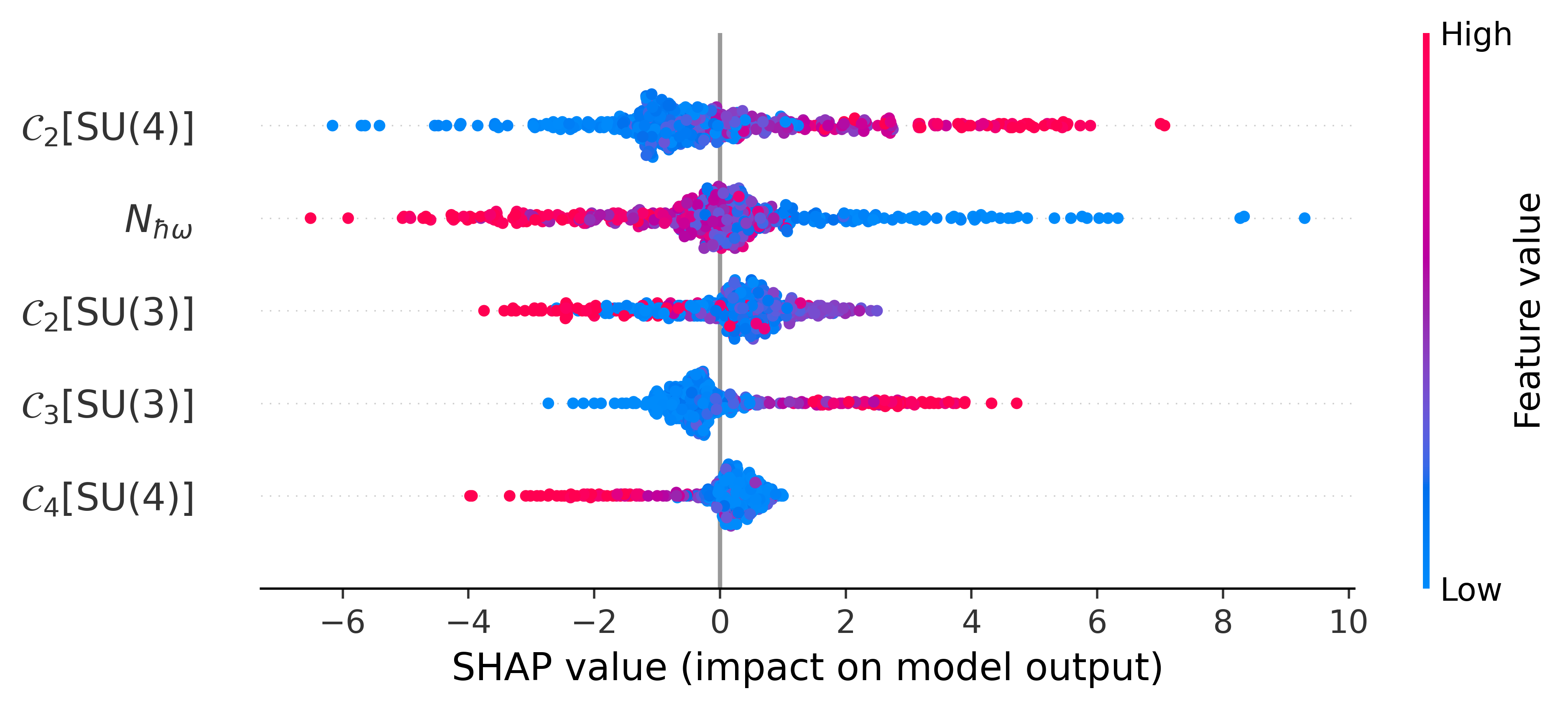}
    \includegraphics[width=0.48\textwidth]{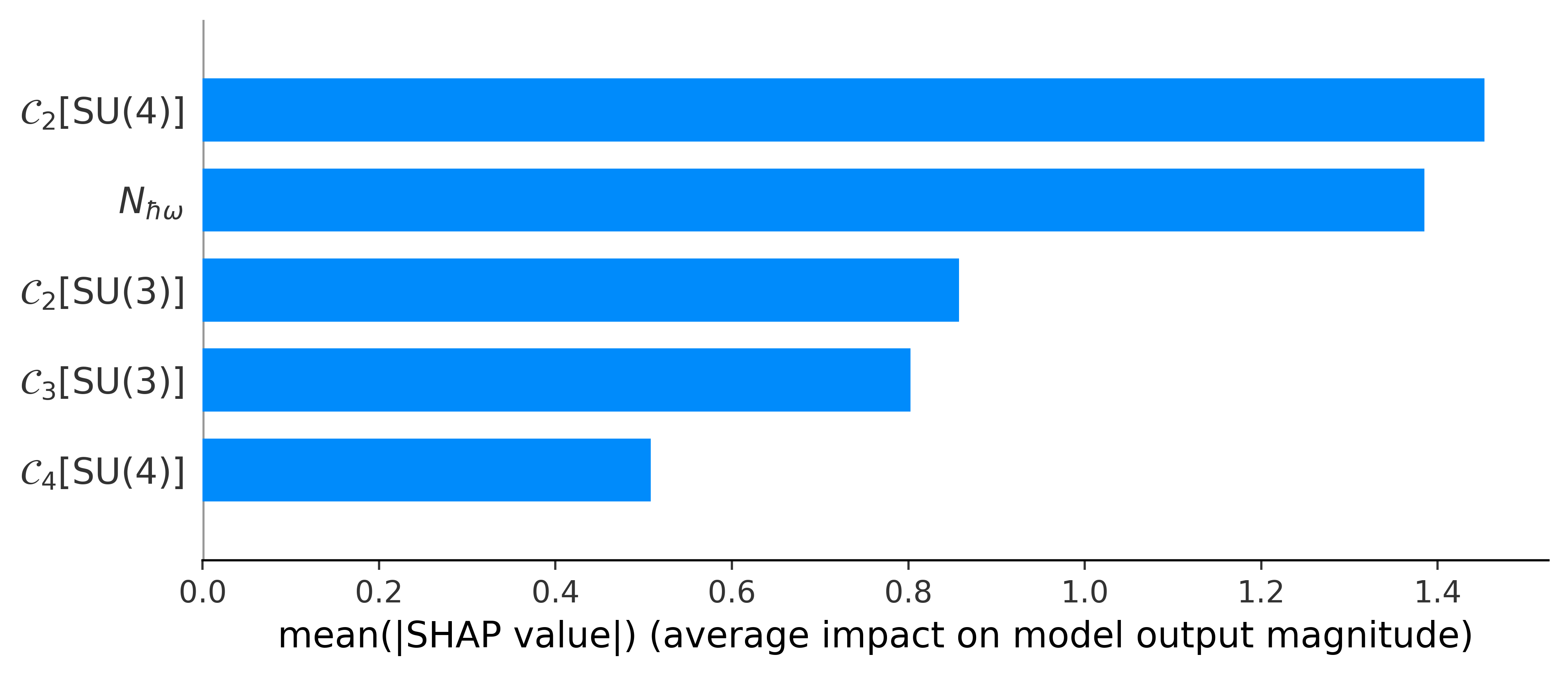}
    \caption{The upper plot is the SHAP summary plot for the GINN at $f=0.8$, evaluated on the test set. Each point represents one nucleus; the horizontal axis gives the SHAP value, i.e., the contribution of a feature to the predicted residual binding energy, $\Delta BE$, relative to the mean prediction. The color encodes the feature value. A positive (negative) SHAP value indicates that the feature pushes the prediction above (below) the mean. Features are ranked from top to bottom by decreasing mean absolute SHAP values, which are shown in the lower plot.}
    \label{fig:shap}
\end{figure}

To rank the influence of the symmetry operators on the predictions, we employ a post-hoc interpretability tool known as Shapley additive explanation (SHAP) values \cite{Lundberg2017} of the GINN (Fig.~\ref{fig:shap}). With the LDM bulk already removed, these values gauge the impact of each operator on the residual binding. The quadratic $\rm SU(4)$ Casimir, $\mathcal{C}_2$, comes out on top, thus Wigner's $\rm SU(4)$ symmetry through its second-order Casimir emerges as the dominant symmetry-derived contributor to binding beyond the bulk properties \cite{Lu2019PLB}. This operator is followed by $N_{\hbar\omega}$ and the $\rm SU(3)$ invariants, which supply the shell structure and quadrupole deformation absent from the LDM. Lastly, the quartic $\mathcal{C}_4[\rm SU(4)]$ trails as a finer, higher-order correction for the binding residual. 
Although $\mathcal{C}_2[\rm SU(4)]$ and $\mathcal{C}_4[\rm SU(4)]$ are both smooth functions of $T_z$ for the leading irreps and therefore co-determined, the dominance of $\mathcal{C}_2[\rm SU(4)]$ is stable and not simply a feature of a particular feature-attribution method; below in Subsec.~\ref{winn-interp} we independently confirm this by the term-by-term contributions read directly off the interpretable WINN, which require no feature-attribution methods.

\begin{figure*}[h!]
    \centering
    \begin{minipage}{0.35\textwidth}\centering
        \includegraphics[width=\linewidth]{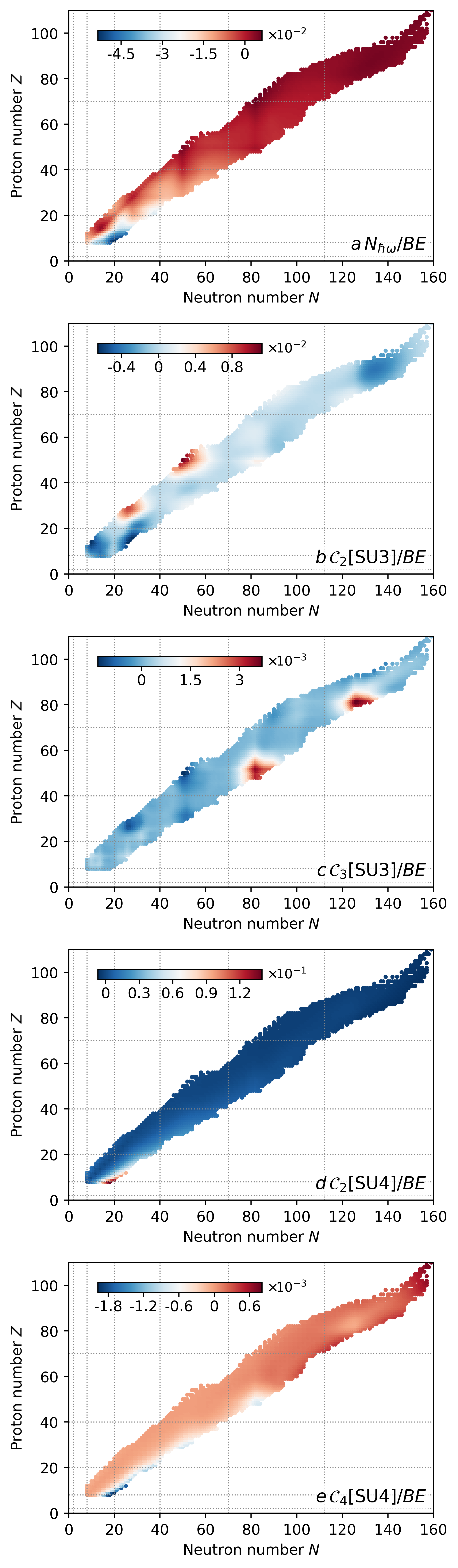}\\[1pt] (a)
    \end{minipage}
    \begin{minipage}{0.35\textwidth}\centering
        \includegraphics[width=\linewidth]{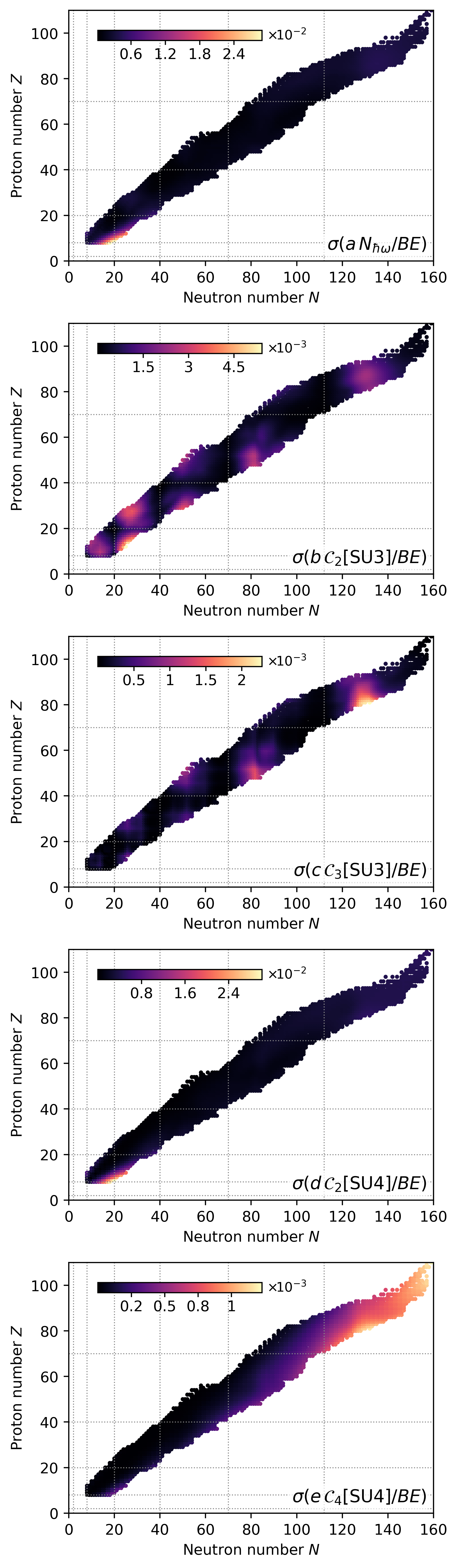}\\[1pt] (b)
    \end{minipage}
    \caption{Term-by-term contribution of the WINN mass formula, Eq.~(\ref{eq:winn}), relative to the total binding energy at $f=0.8$, as an ensemble over 30 trainings (10 initialization seeds $\times$ 3 activation functions: ReLU, PReLU, and Tanh). (a) Ensemble-mean contribution of each term across the nuclear chart; (b) the corresponding standard deviation showing the uncertainty of the decomposition.}
    \label{fig:contributions}
\end{figure*}

\subsection{Interpretation of WINN}
\label{winn-interp}

As argued above, Eq.~(\ref{eq:winn}) is linear in the operators, hence, the trained WINN exposes how each term shapes binding across the nuclear landscape. Crucially, this interpretability does not come at the expense of accuracy---despite its rigid linear form, the WINN is the most accurate of the three as shown in Table~\ref{tab:rmse}---and, because the couplings are learned rather than imposed, the WINN also permits a direct assessment of model-dependent uncertainty in its interpretation, which a post-hoc feature-attribution (like SHAP) does not naturally provide. To quantify it, we retrain the WINN 30 times from different NN initializations at data fraction $f=0.8$ (with ten seeds $\times$ three activation functions, namely ReLU, PReLU, and Tanh), obtaining a validation RMSE of $0.43\pm0.05$~MeV, consistent with Table~\ref{tab:rmse}. Figure \ref{fig:contributions}(a) maps the ensemble-mean contribution of each term in Eq.~(\ref{eq:winn}) relative to the total binding energy, and panel (b) shows its run-to-run ensemble spread as a measure of model-dependent uncertainty in the decomposition.

Consistent with the SHAP analysis of the GINN in Fig. \ref{fig:shap}, $\mathcal{C}_2[\rm SU(4)]$ and $N_{\hbar\omega}$ are the dominant contributors, with peak contributions of roughly $14\pm3\%$ and $6\pm2\%$, respectively. Their dominance is stable across the training ensemble---these two operators are the leading terms in every one of the 30 runs, and the complete ordering is unchanged in 27 of them. The decomposition is reproducible where nuclei are measured, with panel (b) showing a spread below $0.2\%$ of the binding energy across most of the chart; it grows uncertain only toward the data-sparse light and neutron-rich edges. Furthermore, the scales of the contributions from the other operators also reflect the SHAP ranking. 

Beyond this agreement, the WINN reveals physical insights that the SHAP analysis cannot. In particular, the model successfully learns the shell structure encoded in the $\rm SU(3)$ Casimirs. Near shell closures where deformation is low, nuclear binding is largely unaffected by these operators; by contrast, at mid-shell their impacts are enhanced, with $\mathcal{C}_2$ contributing around 1\% and $\mathcal{C}_3$ at the sub-percent level with the sign structure that distinguishes prolate from oblate shapes. 

Furthermore, the even $\rm SU(4)$ Casimirs exhibit an interesting behavior. Specifically, both the quadratic $\mathcal{C}_2$ and the quartic $\mathcal{C}_4$ represent a small fraction of binding energy over most of the chart, however, a pronounced enhancement---reaching beyond 10\% for $\mathcal{C}_2$ with an uncertainty of a few percent---emerges near the neutron dripline, roughly in the region $Z \approx 8$-20. (We observe a similar enhancement when training directly on binding energy instead of the residual.) This enhancement is driven by the learned coupling rather than the operator scale, since the Casimir eigenvalues themselves grow by orders of magnitude toward the superheavy region (Fig.~\ref{fig:features}), where their contribution is instead small. This behavior may be suggestive of a restoration of Wigner's symmetry as the spin-orbit splitting is quenched in neutron-rich species \cite{Lalazissis1998PLB,Fuentes2026PRL,Ding2026PRL}. As Fig. \ref{fig:contributions}(b) illustrates, this edge carries a larger uncertainty, implying that the enhancement is a network-generated hypothesis, and establishing a firmer interpretation is relegated to future work. Note that the degree of SU(4) restoration has been assessed for calcium isotopes \cite{Vogel1993PRC} and more recently for heavy
and superheavy nuclei through Gamow-Teller resonances \cite{Lutostansky2016EPJ}.

\subsection{Testing with $r$-process simulations}
\label{rprocess}

\begin{figure*}[t]
    \centering
    \includegraphics[width=0.7\textwidth]{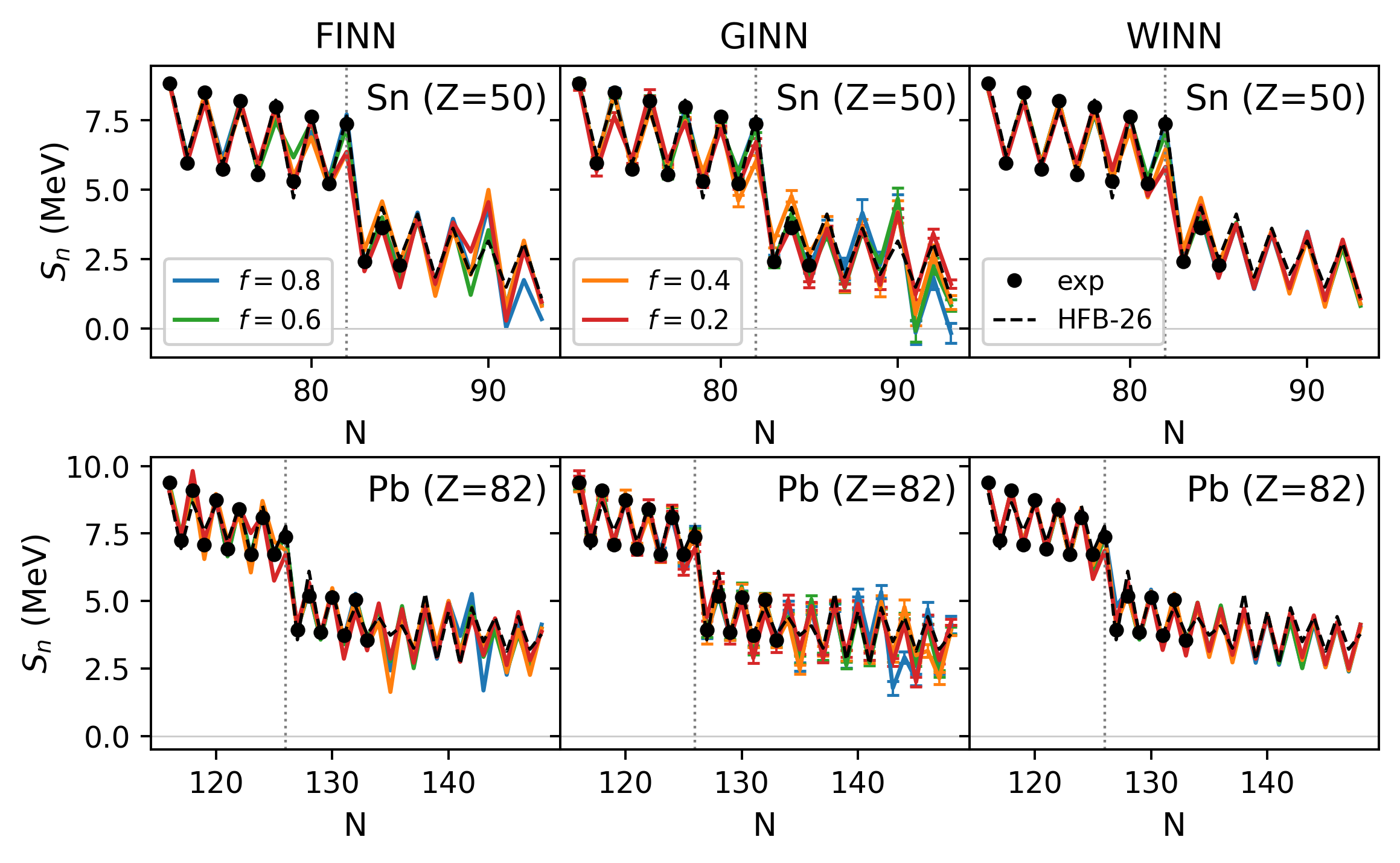}
    \caption{Neutron separation energy $S_n$ along the Sn ($Z=50$) and Pb ($Z=82$) isotopic chains, for FINN, GINN, and WINN at the four training fractions. GINN additionally shows $\pm1\sigma$ bars computed similarly to Ref. \cite{Li2024PLB}. Markers are experiment; dashed lines are HFB-26; vertical dotted lines correspond to the $N=82$ and $N=126$ closures in the top and bottom panels, respectively.}
    \label{fig:sn}
\end{figure*}

\begin{figure*}[t]
    \centering
    \includegraphics[width=0.8\textwidth]{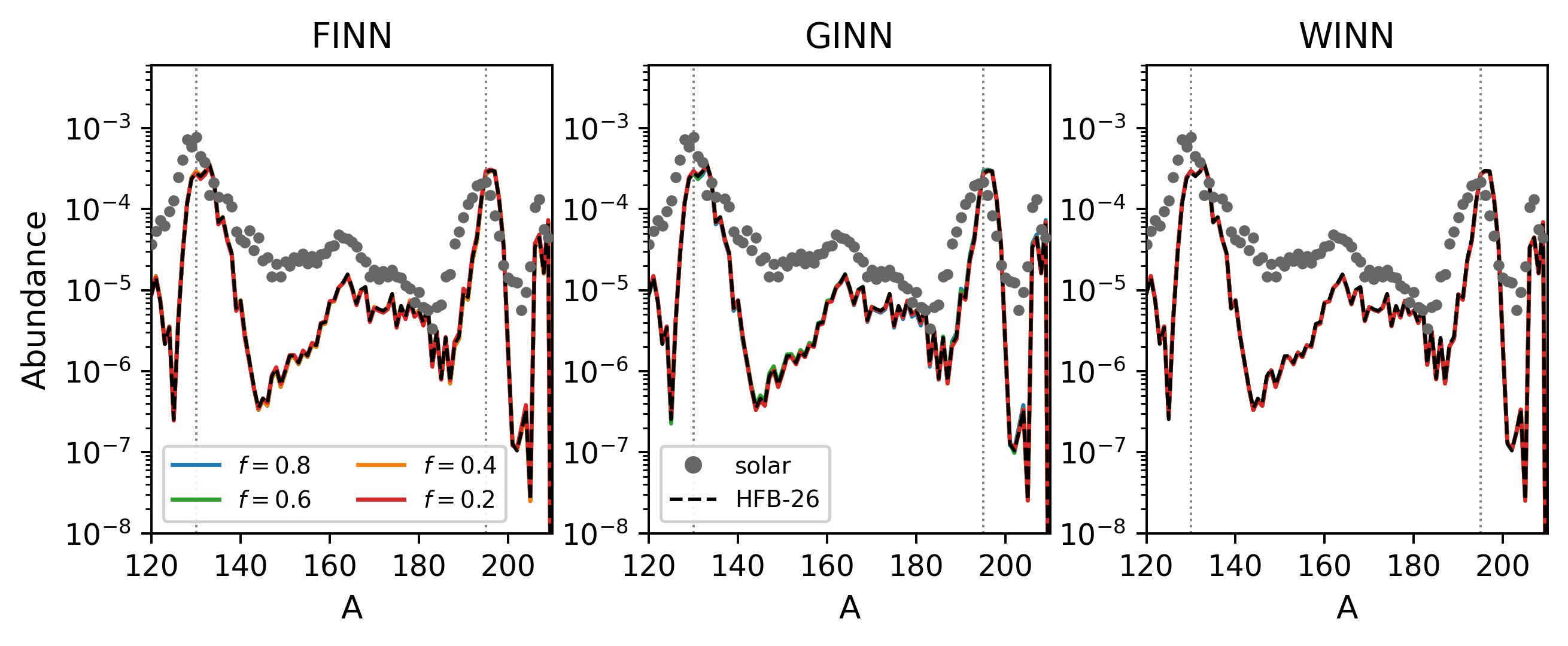}
    \caption{Final $r$-process abundances computed with masses predicted by FINN, GINN, and WINN, each shown for all four training fractions (blue/green/orange/red $= f = 0.8/0.6/0.4/0.2$). Comparison is made with HFB-26 masses (dashed lines) and the solar $r$-process abundances (gray markers). Vertical dotted lines mark the second ($A\approx130$) and third ($A\approx195$) peaks. All curves share the same neutron-star-merger trajectory \cite{Korobkin2012MNRAS}; only the mass table is changed.}
    \label{fig:abundances}
\end{figure*}

Finally, inspired by the work of Li \textit{et al.} \cite{Li2024PLB}, who pioneered the use of machine-learning-predicted masses in $r$-process simulations, we feed our mass tables into a full nuclear reaction network calculation. Using the WinNet library \cite{Reichert2023ApJS} on a representative trajectory of dynamical ejecta from a binary neutron star merger given in Ref. \cite{Korobkin2012MNRAS}, we change only the nuclear mass table---swapping in each model's predictions across the unmeasured region while holding the trajectory, reaction rates, and fission treatment fixed---so that any difference in the yields traces to the masses. Our choice of this trajectory is grounded in the fact that the ejecta is extremely neutron rich, thus the nucleosynthesis products are insensitive to the parameters determining the astrophysical environment and determined mostly by nuclear properties, including masses \cite{Korobkin2012MNRAS}.

For $r$-process applications, neutron separation energies, $S_n$, are particularly important because they largely set the $r$-process path far from stability. Figure~\ref{fig:sn} traces $S_n$ along two chains crossing the $N=82$ and $N=126$ waiting points (Sn and Pb). It can be seen that all models capture the odd-even staggering and the shell drop, and the spread among fractions stays narrow near the closures, opening up only deep in the unmeasured tail. Here again the WINN is the steadiest---its four curves nearly overlap all the way to the dripline---echoing its data efficiency shown in Table~\ref{tab:rmse} and indicating that the architectural constraint plays an important role in stabilizing extrapolation.

The resulting abundances for the three NNs at all four fractions together with the solar $r$-process pattern \cite{Sneden2008ARAA} and that generated by HFB-26 masses are displayed in Fig. \ref{fig:abundances}. By and large, the abundance patterns exhibit similar trends to those in Ref. \cite{Korobkin2012MNRAS}. For every model, the four fraction curves are all but indistinguishable, and every one tracks the HFB-26 results closely, reproducing the second ($A\approx130$) and third ($A\approx195$) peaks. Hence, the abundance pattern is quite insensitive to both the training fraction and the NN mass model---a direct consequence of the well-controlled $S_n$ above. More quantitatively, although the curves appear nearly coincident on the logarithmic scale, they are not exactly identical; specifically, over $120 \le A \le 210$ the four training fractions of a given model differ by at most ${\sim}15\%$ (median ${\sim}5\%$), and the three architectures differ from one another by comparable amounts---up to ${\sim}15\%$ between the FINN and the GINN, and only ${\sim}3\%$ between the FINN and the WINN. These variations are of the same order as the ${\sim}10\%$ spread between the WS3 and HFB-26 abundances themselves, so the pattern is robust to the mass model at the few-percent level. Note that the underproduction near $A \approx 150$ reflects the adopted astrophysical trajectory, rather than the mass models. We also stress that the $r$-process simulations here serve as a probe of the mass tables produced by our NNs rather than an astrophysical study.

\section{Summary and future outlooks}
\label{summary}

We have presented three symmetry-informed NN mass models built on the Casimir operators of Wigner's $\rm SU(4)$ and Elliott's $\rm SU(3)$ symmetries along with the 3DHO number $N_{\hbar\omega}$, trained on \texttt{AME2016} and validated on nuclei first reported in \texttt{AME2020}. Robustness of the mass tables produced by the NNs is tested by benchmarking against HFB-26 in extrapolation and via $r$-process nucleosynthesis. The novelty of our work is three-fold. 

First, the symmetry features carry binding information well beyond the liquid-drop bulk demonstrated through RMSE, and the SHAP analysis ranks the quadratic $\rm SU(4)$ Casimir as the dominant contributor, supporting the relevance of Wigner's symmetry for organizing nuclear binding \cite{Lu2019PLB}. As a physical benchmark, we further verified that the resulting mass tables reproduce the $r$-process abundance pattern robustly across training fractions and in agreement with HFB-26 masses.

Second, the WINN---a domain-informed formula linear in the symmetry operators with only its $(N,Z)$-dependent coupling strengths learned through the NN---reaches the lowest validation RMSE, $0.412$~MeV, competitive with state-of-the-art mass models, and is quite steady against extrapolation. The outperformance of the WINN is evidence for the advantage of embedding the operator structure directly in the architecture rather than supplying it to a black box. We also highlight that the WINN's competitive accuracy matters not only as a benchmark but also as evidence that a compact symmetry basis captures important components of the nuclear binding systematics. Additionally, the analysis of the neutron separation energy along Sn and Pb isotopic chains further shows that the WINN is the most data-efficient among the three.

Third, distinct from previous work on NN mass models, the WINN by design possesses intrinsic interpretability without any post-hoc feature-attribution techniques. Specifically, the WINN coupling strengths illuminate the physics of different mass regions. The $\rm SU(3)$ Casimirs capture deformation-driven shell effects, peaking at mid-shell and suppressed at shell closures. The even $\rm SU(4)$ Casimir contributions are enhanced toward the neutron dripline ($Z \approx$~8-20), which may hint at a restoration of Wigner's symmetry under spin-orbit quenching.

Both the data efficiency and the interpretability of the WINN follow from a single architectural choice, namely constraining the map to be linear in the symmetry operators---a strong inductive bias that shrinks the space of admissible mappings enough to resist overfitting while leaving every term readable across the chart.

It is worth noting that the $\rm SU(3)$ and $\rm SU(4)$ symmetries are assumed in this work to hold exactly across the nuclear landscape, with irrep selection inspired by \textit{ab initio} results. In reality they are broken, a difficulty that has historically been addressed through \textit{pseudo} and \textit{proxy} schemes \cite{Isacker1999PRL,Isacker2023Symmetry,Kota2024PScr,Draayer1984AnnP,Bonatsos2017PRC,Cseh2020PRC}. Our results carry a two-fold implication: (i) the traditional algebraic solutions may improve the NNs through better choices of irreps, and (ii) the NN architecture may itself offer a pathway for resolving the symmetry breaking. More broadly, this work opens a new door in how we engage the emergent symmetries of the nuclear force: where they once organized the structure of individual nuclei, ML now lets them speak to the systematics of the global nuclear landscape. Extending the framework to other observables is a natural next step.

\section*{Acknowledgements}
We thank the \href{https://www.lsu.edu/science/news_events/ai-journal-club.php}{LSU AI journal club} for facilitating collaboration and E. Burns for useful discussions. PD acknowledges support from the Institute of International Education through the Quad Fellowship, and from the National Science Foundation through the Open Questions and Research Tools in Nuclear Astrophysics summer school under Grant No. OISE-1927130 (IReNA). A part of the work of FP was supported by the National Natural Science Foundation of China under Grant No. 12175097. JPD was supported by LSU Sponsored Research Rebate Program and LSU Foundation's Distinguished Research Professorship Program.

\section*{Data availability}

The mass tables produced by our NNs and $r$-process abundances will be deposited in a public database upon publication. The code used to construct the features, train the neural-network models, and reproduce results and figures of this work is publicly available on this \href{https://github.com/phongdang14/Wigner-mass-public}{GitHub repository}. We invite interested researchers and students to contribute and enrich the project.

\appendix

\section{SU(4) Casimir eigenvalues}
\label{app:su4}

Explicit formulas for the Casimir invariants ($\mathcal{C}_2$, $\mathcal{C}_3$, $\mathcal{C}_4$) of the $\rm SU(4)$ group were derived for Wigner's convention of $\rm SU(4)$ labels by Burdet, Maguin and Partensky~\cite{Burdet1968NuovoCim} and by Danos and Gillet~\cite{Danos1972ZP}. Here we adopt an alternative convention of $\rm SU(4)$ irreps by Draayer~\cite{Draayer1970JMP},
\begin{equation}
    P = n_{14}-n_{24}, \quad P' = n_{24} - n_{34}, \quad P'' = n_{34} - n_{44},
\end{equation}
where $[n_{14}, n_{24}, n_{34}, n_{44}]$ is a $\rm U(4)$ irrep~\cite{Dang2024EPJP2} that sums to the number of nucleons, $A=\sum_{i=1}^4 n_{i4}$.

The eigenvalues are derived following the general technique described in Appendix B of Ref.~\cite{Kota2020},  which is applicable to any unitary group. As the Casimirs depend only on the irrep labels, their eigenvalues can be computed in any subgroup chain that is the most convenient, which we choose to be the canonical chain $\rm U(4) \supset U(3) \supset U(2) \supset U(1)$. The $\rm U(4)$ Casimir operators are defined via the canonical generators $E_{ij}$ \cite{Dang2024EPJP2} as
\begin{equation}
\label{eq:U4_Casimir}
    \mathfrak{C}_2 = E_{ij} E_{ji}, \, \mathfrak{C}_3 = E_{ij} E_{jk} E_{ki}, \, \mathfrak{C}_4 = E_{ij} E_{jk} E_{kl} E_{li},
\end{equation}
where summations over repeated indices are implied according to Einstein's convention. Another property of the Casimirs is their independence from any state used to evaluate the eigenvalue; hence, we choose the highest weight state that enables automatic annihilation of half of the terms in Eq.~(\ref{eq:U4_Casimir}). Once the eigenvalues of the $\rm U(4)$ invariants are attained, those of the $\rm SU(4)$ Casimirs are straightforward by a simple conversion from $\rm U(4)$ to $\rm SU(4)$ quantum numbers \cite{Kota2020}. One last step is to invoke a linear mapping that preserves their algebraic properties,
\begin{equation}
    \mathcal{C}_2 = 2 \mathfrak{C}_2, \quad \mathcal{C}_3 = 4(\mathfrak{C}_3 - \mathcal{C}_2), \quad \mathcal{C}_4 = \mathfrak{C}_4 - \mathcal{C}_3,
\end{equation}
and imposes the following symmetry relations under irrep conjugation: $\mathcal{C}_2(PP'P'') = \mathcal{C}_2(P''P'P)$, $\mathcal{C}_3(PP'P'') = -\mathcal{C}_3(P''P'P)$ and $\mathcal{C}_4(PP'P'') = \mathcal{C}_4(P''P'P)$.

Using the matrix elements given in Ref.~\cite{Dang2024EPJP2}, the $\rm SU(4)$ Casimir eigenvalues are
\begin{widetext}
\begin{align}
    \langle \mathcal{C}_2 \rangle =& \tfrac{3}{2}(P^2+P''^2) + 2P'^2 + 2(P+P'') P' + P P'' + 6(P + P'') + 8P', \\
    \langle \mathcal{C}_3 \rangle =& \tfrac{3}{2} (P - P'')[P^2 + P''^2 + 2(P P' + P' P'' + P P'') + 2(3P + 2P' + 3P'' + 4)], \\
    \langle \mathcal{C}_4 \rangle =& \tfrac{1}{64} [21(P^4 + P''^4) + 16 P'^4 + 4P^3 (14 P' + 7 P'' + 42) + 4 P''^3 (14 P' + 7 P + 42) + 32 P'^3 (4 + P + P'') \nonumber \\
                   & + 72 P'^2 (P^2 + P''^2) + 30 P^2 P''^2 + 24 P P' P'' (3 P + 2 P' + 3 P'') + (288 P'^2 + 896 P' + 960)(P + P'') \nonumber \\
                   & + 24 P^2 (16 P' + 9 P'') + 24 P''^2 (16 P' + 9 P) + 384 P P' P'' + 576 (P^2 + P''^2) + 512 P'^2 + 640 P P'' + 1024 P' ].
\end{align}
\end{widetext}


\section{SU(3) Casimir eigenvalues}
\label{app:su3}

We follow the convention for the quadratic ($\mathcal{C}_2$) and cubic ($\mathcal{C}_3$) $\rm SU(3)$ Casimir eigenvalues of Draayer and Rosensteel~\cite{Draayer1985NPA}, which can be derived analogously to the $\rm SU(4)$ case,
\begin{align}
\langle \mathcal{C}_2 \rangle &= \lambda^2 + \mu^2 + \lambda\mu + 3\lambda + 3\mu, \\
\langle \mathcal{C}_3 \rangle &= \tfrac{1}{9} (\lambda -\mu)[2\lambda^2 + 2\mu^2 + 5\lambda\mu + 9(\lambda + \mu + 1)],
\end{align}
where $\lambda$ and $\mu$ are labels of an $\rm SU(3)$ irrep. Under irrep conjugation, $\mathcal{C}_2(\lambda\mu) = \mathcal{C}_2(\mu\lambda)$ and $\mathcal{C}_3(\lambda\mu) = -\mathcal{C}_3(\mu\lambda)$ hold.

\bibliographystyle{apsrev4-2}
\bibliography{References}

\end{document}